# Quality of Service Improvement for High-Speed Railway Communications

Zhou Yuzhe, Ai Bo*

State Key Laboratory of Rail Traffic Control and Safety, Beijing Jiaotong University, No.3 Shang Yuan Cun Haidian District, Beijing, 100044, P. R. China.

**Abstract**: With the fast development of high-speed railways, a call for fulfilling the notion of communication at "anytime, anywhere" for high-speed train passengers in the Train Operating Control System is on the way. In order to make a realization of that, new railway wireless communication networks are needed. The most promising one is the Long Term Evolution for Railway which will provide broadband access, fast handover, and reliable communication for high mobility users. However, with the increase of speed, the system is subjected to high bit error rate, Doppler frequency shift and handover failure just like other system does. This paper is trying to solve these problems by employing MIMO technique. Specifically, the goal is to provide higher data rate, higher reliability, less delay, and other relative quality of services for passengers. MIMO performance analysis, resource allocation, and access control for handover and various services in a two-hop model are proposed in this paper. Analytical results and simulation results show that the proposed model and schemes perform well in improving the system performances.

## I. INTRODUCTION

With the rapid construction of high speed railway (HSR) and dedicated passenger railway lines, Global System for Mobile Communication for Railway (GSM-R) System [1] and Long Term Evolution for Railway (LTE-R), as railway wireless communication networks, are used in the China Train Operating Control System Level 3 (CTCS-3) to fulfill the high speed railway reliability requirements of signal transmitting. However, as the fast development of the Internet, high-speed train passengers want to access the Internet at "anytime, anywhere". This unavoidable trend puts forward much higher requirements on high-speed railway communication services. In order to keep up with this unprecedented growth of service demands, current researches mainly focus on providing wireless network services, i.e. broadband access, fast handover, and reliable communication for high mobility users. Based on the nature of the communication system, trade-offs should be made. And efficiency, reliability, and real-time are the most basic and important aspects amid these trade-offs. To this end, try to provide broadband, real-time, and reliable communication together for high speed users is the main task for researchers.

With the speed increases, the system will subject to high bit error rate, Doppler frequency shift and handover failure. High bit error rate (BER) may cause signaling error, retransmission, and energy waste. The Doppler frequency shift will degrade the quality of the communication link. And handover failure may result in call drops. Through space-time or space-frequency block coding techniques [2, 3], Multiple Input Multiple Output (MIMO) system can provide diversity gain, coding gain, and array gain, respectively. Thus, by converting and utilizing the multi-path effect, this "disadvantage" can reduce the BER. Many works [4-6] can be found to properly model MIMO channels in the high-speed scenario. However, as the train is high-speed moving, channel state information (CSI) may be inaccurate and then robust transceiver design [7, 8] is an important issue. Besides, Doppler frequency shift will cause inter-channel-interference (ICI) which decreases the link data rate in systems that applying Orthogonal Frequency Division Multiplexing (OFDM) technology. The authors in [9] proposed a new vehicular MIMO channel model and deals with the ICI impact of the proposed model. The proposed scheme has better ICI mitigation performance and much less implementation complexity. Since the communication system of HSR is also a kind of cellular networks, it assigns a number of radio channels according to the power constraints and spectrum availability. When the mobile user enters another cell, and the user's ongoing service has not finished yet, the network must handover [10] the service to the target cell around the cell boundary without user's awareness of handover and without reducing the service quality too much. A critical question raised by handovers is which location is the best for handover [11].

Providing Quality of Service (QoS) in wireless networks with limited resources is a challenging problem, especially when users are moving at a very high speed. Network service access control [12] technique is a fundamental mechanism to provide good QoS. It determines the accessibility to the network on account of the resource availability in order to avoid network congestion and service degradation for online users. A new service request will be accepted if and only if the remaining resources are enough to ensure the QoS requirements of the new service without impairing the QoS of already accepted services. Many works had made considerations about application priority and higher layer QoS [13] according to the network QoS architectures. The authors in [14] give a detailed analysis of resource allocation for high speed train communications. On-demand data service is mainly considered and a pre-downloading algorithm is proposed and tested on a real high-speed train schedule. However, the effects of physical layer transmission, the handover issue, and different service priority problem are not covered.

In this paper, we first analyze the system QoS requirements and MIMO performance for HSR. Then a key parameter is obtained via solving a multidimensional resource allocation optimization problem whose object is minimizing the total transmit power together with satisfying the users' QoS requirements. Next, a dynamic access control scheme is proposed so as to trade off handover and new services when the system's resources are limited. Additionally, two modified schemes which take into account of the relative priority of the application types and the application overhead. Finally, we give all the proposed approaches' simulation results with comparisons.

The paper is organized as follows. Firstly, we describe the high speed train communication model with MIMO and some analytical methods to achieve QoS requirements in Section II. In Section III, we derive the resource allocation problem and apply the solution to modify the access control scheme for handover and new calls for different services. Simulation results are shown in Section IV. Section V concludes this paper.

## II. SYSTEM MODEL

### 2.1 Networks architecture

As illustrated in Figure 1, the high-speed railway communication network comprises building baseband unit (BBU) and radio remote unit (RRU). For the sake of reliable communications between RRU and the high speed train (HST), two mobile relay units (MRUs) are equipped on the top of the former and the rear of the HST, respectively. They can operate cooperatively or independently according to different scenarios. Meanwhile, each train carriage is employed with a repeater which is connected to the MRUs. Users can access networks by these distributive repeaters. Thus a two-hop model rises, i.e. the first hop refers to the RRUs to the MRUs, and the second hop refers to MRUs/repeaters to the users. In this model, the MRU is employed as a kind of relay station in order to overcome the large penetration loss. We will consider the access problem of each hop in Section 3, respectively. Moreover, MIMO antennas are adopted in the communication system of HSR, which means the RRUs and the MRUs can have multiple antennas.

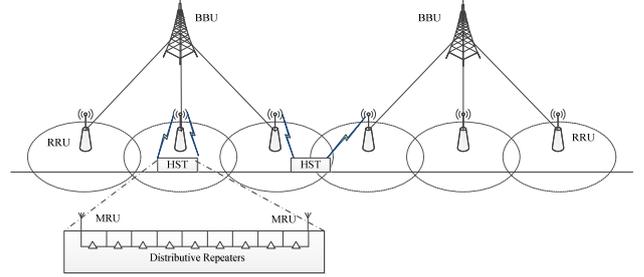

**Fig. 1 HSR MIMO system**

In our proposed HSR system model, for example, each RRU has $N_t$=2 antennas, and each MRU has $N_r$=2 antennas. Joint transmission is performed at 2 adjacent RRUs when handover occurs. According to Diversity-Multiplex-Tradeoff [15], we argue that in the handover scenario, diversity gain should be maximized to minimize the probability of outage and call dropping. While in the non-handover scenario, multiplex gain should be large enough to ensure a higher data rate. So in this paper, we assume that $2\times4$ MIMO will provide a multiplex gain of 2 in a non-handover scenario, and $2\times2$ MIMO will only provide a diversity gain of 4 in the handover scenario.

### 2.2 MIMO system analysis

#### 2.2.1 SVD Method

Assuming that the RRUs and the MRUs are employed with $N_t$ transmit antennas and $N_r$ receive antennas, respectively. The received signal can be expressed as:

$$\mathbf{y} = \mathbf{Hx} + \mathbf{n} \quad (1)$$

where $\mathbf{x}$ is $N_t$-dimensional transmitted signal vector, $\mathbf{H}$ is $N_r \times N_t$ channel matrix, $\mathbf{n}$ is $N_r$-dimensional noise vector, $\mathbf{y}$ is $N_r$-dimensional received signal vector. $I$ is the rank of $\mathbf{H}$. According to the singular value decomposition (SVD) theorem, channel matrix $\mathbf{H}$ can be decomposed into the following expression:

$$\mathbf{H} = \mathbf{U}\begin{bmatrix} \mathbf{D} & \mathbf{0} \\ \mathbf{0} & \mathbf{0} \end{bmatrix}\mathbf{V}^H \quad (2)$$

where $\mathbf{U}$ is $N_r \times N_r$ unitary matrix, and $\mathbf{D}=diag(\sigma_1,\sigma_2,...\sigma_I)$ is $I \times I$ diagonal matrix, the diagonal elements are the singular values of $\mathbf{H}$, and $\mathbf{V}$ is $N_t \times N_t$ unitary matrix. We define three matrix transformations as follows: $\tilde{\mathbf{y}} = \mathbf{U}^H\mathbf{y}$, $\tilde{\mathbf{x}} = \mathbf{U}^H\mathbf{x}$, and $\tilde{\mathbf{n}} = \mathbf{U}^H\mathbf{n}$. Then to apply pre-processing, expression in (1) can be substituted by:

$$\tilde{\mathbf{y}} = \begin{bmatrix} \mathbf{D} & \mathbf{0} \\ \mathbf{0} & \mathbf{0} \end{bmatrix} \tilde{\mathbf{x}} + \tilde{\mathbf{n}}. \quad (3)$$

An equivalent scalar expression is as follows:

$$\begin{cases} \tilde{y}_i = \sigma_i \tilde{x}_i + \tilde{n}_i, & i = 1, 2, \ldots, I. \\ \tilde{y}_i = \tilde{n}_i, & i = I+1, I+2, \ldots, N_r. \end{cases} \quad (4)$$

For the channel matrix whose size is $N_r \times N_t$, the number and the value of the singulars $\sigma_i, i=1,\ldots,n_{nim} = \min(N_r, N_t)$ have a close relation with the performance. According to [15], the parameter $I$ is the number of non-zero singulars and it is the number of spatial degrees of freedom per second per hertz. Note that the matrix is well-conditioned if the condition number $\max_i \sigma_i / \min_i \sigma_i$ is close to 1. At high SNR regime, well-conditioned $\mathbf{H}$ facilitate communications. While at low SNR regime, the MIMO channel provides a power gain of $\max_i \sigma_i^2$.

**2.2.2 Small Scale Loss**
In this paper, we assume that the channel state information is available at both transmitter and receiver. Note that this information can be obtained by exploiting uplink-downlink channel reciprocity in a TDD system or through downlink channel estimation and feedback in an FDD system. However, in practical systems, conventional SVD-based transmission usually suffers dramatical degradation from the time variations of the channel. The authors in [16] proposed a transmission strategy against high-speed mobility in TDD MIMO systems. The results show that the proposed scheme can efficiently reduce the effect of the CSI impairment at the transmitter caused by channel variation. Therefore, under a good CSI assumption, the MIMO channels can be represented into $I$ parallel single-input single-output (SISO) channels that do not interfere with each other. The channel gain of each SISO channel is $\sigma_i$ ($i = 1, 2, \ldots, I$) in Eq. (4). We also assume that the equivalent SISO channel gains obey Nakagami-m distributions.

**2.2.3 Doppler Shift**
For the HSR communication scenario, the HST moves at a speed typically above 300 km/h, this will engage a serious Doppler shift. Thus, the orthogonality of different OFDM sub-carriers is destroyed, which will result in inter-carrier interference. According to [17], the total ICI power on the $n$-th sub-carrier can be expressed as:

$$ICI_n = \frac{(T_s f_d)^2}{2} \sum_{j=1, j \neq n}^{N} \frac{1}{(j-n)^2} \quad (5)$$

where $T_s$ is the period of OFDM symbol, and $f_d$ is the maximum Doppler shift which can be calculated by:

$$f_d = \frac{v}{c} f_c$$

where $v$ is the speed of the HST, and $c$ is the speed of electromagnetic wave, and $f_c$ is the central carrier frequency.

Taking the above analysis into consideration, the signal-to-interference plus noise ratio (SINR) can be described as:

$$SINR = \frac{P_D}{P_I + P_N} = \frac{\sigma^2 P}{ICI + n_0 W} \quad (6)$$

where $P$ is the transmit power, and $n_0$ represents the single-sided noise power spectral density, and $W$ indicates spectrum bandwidth. Note that the SINR term is critical in the following analysis.

**2.2.4 QoS Requirements**
The system QoS requirements, including data rate and BER, partially depend on the modulation and coding scheme (MCS). In the following part, we only consider these two QoS parameters. According to 3GPP RAN1 [18], Table 1 shows the required received SINRs for MCSs at a desired BER of $10^{-5}$.

**Table 1 Required SINRs for MCSs for BER=$10^{-5}$**

| MCS Scheme | SINR(dB) | Data Rate (Mbps/MRU) |
|---|---|---|
| QPSK, R=1/2 | 2.1 | 18.637 |
| QPSK, R=3/4 | 3.0 | 27.956 |
| QPSK, R=7/8 | 4.7 | 32.615 |
| 16QAM, R=1/2 | 6.8 | 37.274 |
| 16QAM, R=3/4 | 7.0 | 55.911 |
| 64QAM, R=3/4 | 10.6 | 83.867 |

From Table 1, we can see that in order to guarantee a peak data rate of 100Mbps, the link from RRUs to MRUs should apply MCS of 16QAM and R=3/4 to both of the two MRUs cooperatively. Then the wireless link on the first hop of the system needs an SINR of no less than 7dB to achieve a total peak data rate of 2×55.911=111.822Mbps. However, when the HST is in a handover zone, only half of the data rate can be achieved. Therefore, the MCS should be changed accordingly, i.e. QPSK and R=3/4 for an SINR of no less than 3 dB.

## III. PROBLEM FORMULATION

In order to guarantee the required SINR performance, one approach is to use a large fixed transmitting power. However, for the sake of Green Communication, large transmitting power is not a good solution. Another approach is to employ power control to minimize the total power consumption. We consider a more general problem formulation, i.e. multidimensional resource allocation. We follow the same assumptions in [19], there are $N$ sub-carriers that can be assigned to $K$ users from $I$ antennas during $T$ time slots. For the $k$-th user, in order to guarantee the QoS demands, the lowest data rate is $R_k$ bits/OFDM symbol, and the target BER is $P_k$. Note that different applications require different $R_k$ and $P_k$, this represents different resource requirements for each user $k$. Moreover, let $b_{k,n,i,t}$ represents the number of bits that are modulated on the $n$-th subcarrier from the $i$-th antenna at the $t$-th time slot of the $k$-th user. And variable $\delta_{k,n,i,t} \in \{0,1\}$ represents the state of resource allocation, i.e. 1 means allocate and 0 means not allocate.

For the purpose of acquiring the QoS demands, allocated transmitted power of the four dimensions is given by:

$$p_{k,n,i,t} = \delta_{k,n,i,t} \frac{SINR_{req}(ICI_n + n_0 W)}{\sigma^2_{k,n,i,t}} \quad (7)$$

where $SINR_{req}$ is the required received SINR, and $\sigma_{k,n,i,t}$ denotes channel gain. The system applies M-QAM modulation scheme. According to [20], the BER of M-QAM is expressed as:

$$P_{b,M-QAM} \approx \frac{4\left(1-\frac{1}{\sqrt{M}}\right)}{\log_2 M} Q\left(\sqrt{\frac{3\log_2 M}{M-1}\frac{E_b}{N_0}}\right), \quad M \geq 2. \quad (8)$$

After some simple calculations, an upper bond can be set to:

$$P_{b,M-QAM} \leq 4Q\left(\sqrt{\frac{3\log_2 M}{M-1}\frac{E_b}{N_0}}\right)$$
$$\leq 4Q\left(\sqrt{\frac{3}{M-1}SNR}\right) \quad (9)$$
$$\leq 4Q\left(\sqrt{\frac{3}{M-1}SINR}\right).$$

Thus, we can rewrite our $P_k$ into the form below:

$$P_k = 4Q\left(\sqrt{\frac{3SINR_{req}}{2^{b_{k,n,i,t}}-1}}\right) \quad (10)$$

where $Q(x) = \frac{1}{\sqrt{2\pi}}\int_x^\infty e^{-\frac{t^2}{2}} dt$. Note that a fixed $P_k$ of $10^{-5}$ can be achieved because the required SINR is gained according to Table 1 through the good performance of MIMO techniques. However, in this problem we reverse the calculation, if we fix a BER of $10^{-5}$ and in order to gain a required SINR, the minimum transmitting power is desired.

### 3.1 Multidimensional resource allocation

Several dynamic resource allocation schemes are presented in [21, 22]. The authors in [19], formulated a multidimensional resource allocation problem whose object is to minimize the total power to be transmitted and set each user's QoS requirements as constraints. Note that this problem is dealt in the first hop of the system, i.e. the MRUs are seen as users. Furthermore, for the $n$-th subcarrier, it can only be allocated to one user from one antenna at the $t$-th time slot. This is an optimization problem that can be formulated as follows:

$$\min_{\delta,b} \sum_{k=1}^{K}\sum_{n=1}^{N}\sum_{i=1}^{I}\sum_{t=1}^{T} p_{k,n,i,t}$$

$$s.t. \begin{cases} p_{k,n,i,t} = \delta_{k,n,i,t}\dfrac{(ICI_n + n_0 W)(2^{b_{k,n,i,t}}-1)}{3\sigma^2_{k,n,i,t}}\left(Q^{-1}\left(\dfrac{P_k}{4}\right)\right)^2 & (11) \\ \sum_{n=1}^{N}\sum_{i=1}^{I}\sum_{t=1}^{T}\delta_{k,n,i,t}b_{k,n,i,t} \geq R_k, \quad \forall k \\ b_{k,n,i,t} \in \{0,2,4,6\}, \quad \forall k,n,i,t \\ \delta_{k,n,i,t} \in \{0,1\}, \quad \forall k,n,i,t \\ \sum_{k=1}^{K}\sum_{i=1}^{I}\delta_{k,n,i,t} = 1, \quad \forall n,t \end{cases}$$

In this paper, the object of the above procedure is to obtain the number of bits to be modulated, i.e. presented as $b_{k,n,i,t}$. The values 0, 2, 4, and 6 denote no transmission, QPSK, 16QAM, and 64QAM modulation schemes, respectively. Note that this parameter can be used in our proposed access control scheme.

### 3.2 Application access schemes

When the HST is in the handover zone, the rear MRU is still communicating with the source RRU and the front MRU is establishing links with the target RRU. In this scenario, only 50 Mbps data rate is achieved together with the same BER of the non-handover scenario. The system bandwidth will shrink to half compared with the non-handover scenario, and the other half bandwidth can be reserved for hand the ongoing call over via the front MRU with lower MSC schemes according to Table 1. In our scheme, we consider three kinds of services -- data, voice, and video, respectively. Note that this problem is dealt in the second hop of the system, so the total resource is obtained by MRUs and will be assigned to the users inside the train by the distributive repeaters. The main metric for evaluating the access capability of the schemes is access ratio. We mark AR as the abbreviation of access ratio. And the definition of AR is:

$$AR = \begin{cases} \dfrac{Total\_accepted\_calls}{Total\_calls}, & Total\_calls \neq 0 \\ 0, & Total\_calls = 0 \end{cases} \quad (12)$$

where the $Total\_accepted\_calls$ of an application means the total number of calls which is accepted by MRUs/RRUs in the simulation time. The $Total\_calls$ of an application means the total number of calls which is generated in the simulation time. Note the calls contain both the handover calls and the new calls. The AR is in the range of [0, 1].

### 3.2.1 Scheme considering handover and new services

Since dropping an on-going service during a handover will be more annoying than blocking a new service, to this end, in the access control scheme, the handover calls should be given higher priority over the new calls. Assuming the access control threshold for handover call and new call are $I_{th}^{HO}$ and $I_{th}^{NEW}$, respectively. $\Delta I_{th} = I_{th}^{HO} - I_{th}^{NEW}$ is the size of the resource reservation. Note that if $\Delta I_{th}$ always remain the same, this method is called fixed resource reservation, but it has the following problems: If the resource reservation, say $C_{total}$ is too large, although it is possible to ensure the handover success rate, but it can also greatly reduce the access ratio and the system resource utilization regard to new calls. In addition, if the resource reservation $C_{total}$ is too small, it will be insufficient to guarantee handover success rate. A better way is to apply adaptive priority access control scheme which dynamically adjusts the reserved size of the resource. We define the resource reservation factor $\beta$ as follows:

$$\beta = \begin{cases} \dfrac{x}{D_1}, & x < D_1 \\ 1, & D_1 \leq x \leq D_2 \\ \dfrac{D_{overlap} - x}{D_{overlap} - D_2}, & x > D_2 \end{cases} \quad (13)$$

where $D_{overlap}$ is the size of the overlapping area. Because a handover will and should trigger with a very high probability in the central area of the overlapping zone, we denote these boundaries by $D_1$ and $D_2$, so we let $\beta$ higher when the HST is in the central area of the overlapping zone.

Because handover call has higher priority, it can compete for all the resources, so we make sure the resource reservation factor for handover is large enough. Handover call admission control threshold $I_{th}^{HO}$ is determined by the initial network planning, while new call admission control threshold is given by $(1-\beta)I_{th}^{HO}$, then the flow chart of the adaptive priority access control scheme is shown in Figure 2, where $C_{current}$ denotes the current application resource request; and superscripts $H$ and $N$ denote the handover and new calls, respectively.

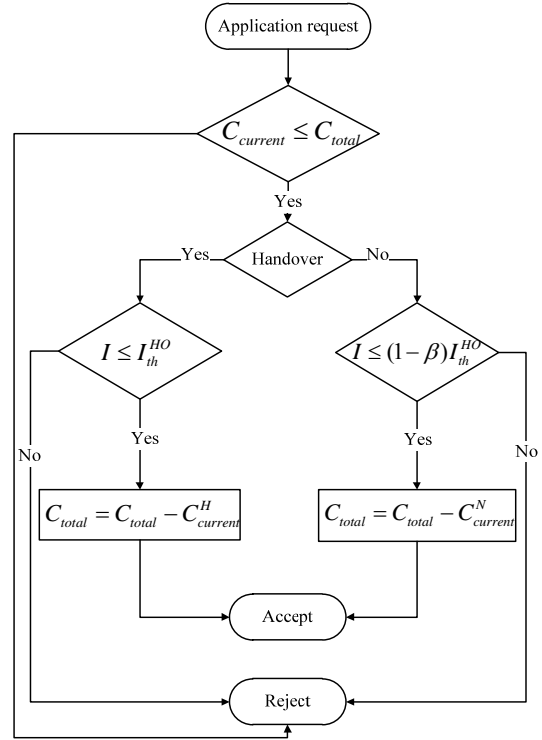

**Fig. 2 Access control scheme**

### 3.2.2 Scheme considering application priority

Different users request to access services at the same time is quite popular in the HSR communication scenario. There are usually hundreds of passengers in the HST. There is a high probability that some passengers request some services simultaneously. Besides, when one HST or two (in opposite directions) at the same time are moving in a cell, ongoing services must be handover to this cell simultaneously as well. The above two types of services are called new service and handover service, respectively. However, due to the limited resource, some of these requests may be rejected. So how to allocate the limited resources to the user services needs careful considerations [23, 24]. Generally speaking, three applications: voice, data, and video are commonly used by users. According to the QoS architecture, they should have the following priority relationship:

$$P_{voice}^H > P_{video}^H > P_{data}^H > P_{voice}^N > P_{video}^N > P_{data}^N \quad (14)$$

where $P_{voice}$, $P_{video}$, and $P_{data}$ are the priority of voice service, data service, and video service, respectively; and superscript $H$ and $N$ denote handover and new service, respectively.

The idea of the access control scheme with respect to application priority is described as follows. When more than one service requests to access the network at the same time or time slot, handover services will be considered prioritized as the previous part described. Different handover services will be determined whether can be accepted or not based on their application priorities. The decision scheme is that in $k$-th decision loop, check whether the criteria $C_{total}^{(k+1)} = C_{total}^{(k)} - C_{current}^{H\,or\,N} \geq 0$ is satisfied, $C_{total}^{(k)}$ is the total remain resource in $k$-th loop, and $C_{current}^{H\,or\,N}$ means the size of resource will be occupied by one of the handover or new call requests, note that each decision loop subtracts only one term in an order defined in (14), and accepts the current request if the criteria is satisfied and go on $k+1$-th loop. After all handover services have been processed, different new services will be processed similarly as the handover ones.

### 3.2.3 Scheme Considering Application Priority and Overhead

The above principles of the access control schemes only take the user data into consideration. However, there is a large amount of overheads when the user is requesting a service, and the overhead will definitely take up a certain amount of resources to be allocated to users. When the speed of the HST is $v$, parameter $b$ can be acquired according to Section 3.1. There is a relationship of $P_s = 1-(1-P_B)^b$ [25] between BER $P_B$ and symbol error rate $P_s$ provided that $b$ is even number. Assuming one packet contains $N_s$ symbols, so the number of error symbols is $N_{error} = N_s P_s$. We assume that the number of check symbols required is $N_{check} = \alpha N_{error}, \alpha > 1$. Since the additional overhead for a packet includes header and check symbols, the additional overhead in our proposed scheme will vary according to parameter $b$. The overhead per bit is expressed as:

$$O_p = \begin{cases} \dfrac{H_b + bN_{check}}{bN_s}, & b \neq 0 \\ 0, & b = 0 \end{cases} \quad (15)$$

where $H_b$ is the header. The $O_p$ will grow with the increase of speed, which will result in the decrease of effective data rate. With regard to a user who currently takes up $C_{current}$ resources, in this case, more are needed to provide the QoS requirement. Therefore, the final resources the user needed is presented as:

$$C'_{current} = (1+O_p)C_{current}. \quad (16)$$

In addition, the access control scheme with application priority should be modified with regard to overhead concern as $C_{total}^{(k+1)} = C_{total}^{(k)} - C'_{current} \geq 0$, the similar explanation as previous describes.

## IV. SIMULATION RESULTS

### 4.1 Simulation settings

We assume that each RRU has $N_t = 2$ antennas, and each MRU has $N_r = 2$ antennas. MCS is applied according to Table 1. Other parameters are shown in Table 2. Three applications---voice, data, and video, are simulated together. The required data rates/bandwidths are 64, 128, and 512Kbps, respectively. And each call arrival is modeled as a Poisson process with arrival rates of 2, 0.5, and 0.1 per second, respectively. The average service time follows exponential distributions with mean 60, 300, and 600 seconds, respectively.

**Table 2 Simulation parameters**

| Parameter | Value |
|---|---|
| Bandwidth | 20 MHz |
| Carrier Frequency | 2.3 GHz |
| Number of Sub-carrier | 1024 |
| Transmit Power | 46 dBm |
| RRU Cell Radius | 4 km |
| Overlapping Distance | 1 km |
| Symbols per Packet | 2048 |

### 4.2 MIMO performance in HSR

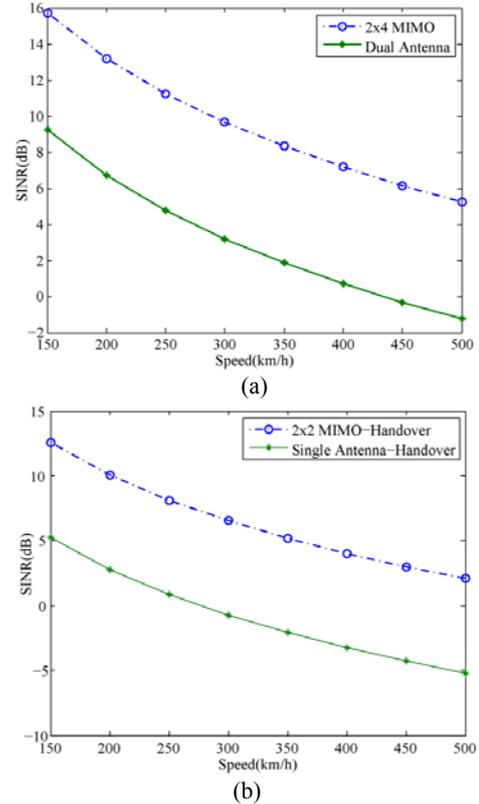

(a)

(b)

**Fig. 3 Proposed Model Performance**

Figure 3 (a) shows the SINR performance of the $2\times 4$ MIMO scheme and the dual antenna scheme which can be seen as a $1\times 2$ MIMO scheme both in non-handover scenarios. As we can see, in order to ensure a requested SINR of 7 dB, the proposed $2\times 4$ MIMO can guarantee train speed below 400 km/h, while the dual antenna scheme can guarantee only a maximum of 250 km/h. In fact, we can adopt different MCS schemes when the HST is at different speed according to Table 1, thus a higher data rate can be achieved. Figure 3 (b) shows the SINR performance of the 2×2 MIMO scheme and the single antenna scheme which is a point-to-point system in a handover scenario. In order to ensure a requested SINR of 3 dB, the proposed 2×2 MIMO scheme can guarantee train speed below 400km/h, while the single antenna scheme cannot guarantee this SINR. Thus, the dual antenna scheme is good at non-handover scenario. However, it will not meet the QoS requirements when the system is reduced to a point-to-point system in a handover scenario, which will cause call drops. The simulation results show that our proposed MIMO scheme for high speed rails will meet the SINR request. Therefore it can provide better QoS for train communications.

### 4.3 Solving the Optimization Problem

We use Matlab to simulate the system and use Lingo to solve the multidimensional optimization problem. We reformulate the original problem to a similar one by assuming that the weights of the object function are the same, i.e. all equal to 1. After we find a solution, we sum the overall item by using the transmitted power as weights to the solution. This will yield a suboptimal solution with much less computational complexity.

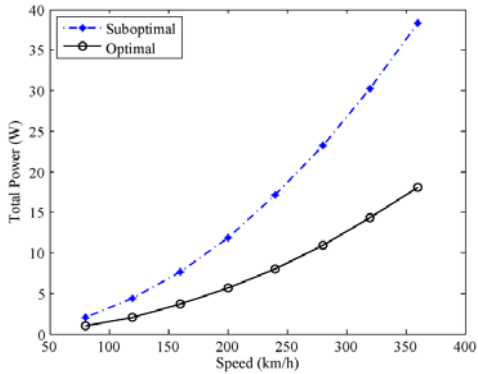

**Fig. 4 Comparison of the solutions**

Figure 4 shows the solutions of this problem. In order to guarantee the required QoS, say SINR, the total transmitted power from one antenna of the MIMO system is increasing with regard to the speed. And note that the maximum power used to transmit is less than the simulation value 40W. So, we can conclude that by using multidimensional power control, our proposed MIMO scheme can guarantee the requested QoS with less power consumption than the fixed one. Note that the suboptimal solution is larger than the optimal solution and the gap is increasing with the speed. Although the suboptimal solution is a kind of wasting energy, it reduces a great amount of calculation compared with the optimal solution. This is critical when the HST is at higher speeds, since the HST will have much less time to pass a cell and handover. For the real time resource allocation, we'd better apply the suboptimal method to provide the real time property of the system.

### 4.4 Access Control Scheme

Figure 5 (a) shows the access ratios of handover calls and new calls of different applications in the handover zone. When the speed increases, the new-call access ratio of all applications remain a low level, while the handover-call access ratios of all applications tend to increase and can be much higher than the new calls. This is because when the speed is higher, the resource reservation factor makes it more prone to access handover calls and less likely to access new calls. Additional comparison of handover calls and new calls in the non-handover zone is shown in Figure 5 (b). The handover of voice service is given particular care, since the ratio is higher than the new voice calls in the non-handover zone. However, access ratios of data and video services have crossing points under these two scenarios, which means the resource reservation factor can be more efficient in higher mobility. So with the help of dynamic resource reservation scheme, QoS of handover calls is guaranteed prior to new calls.

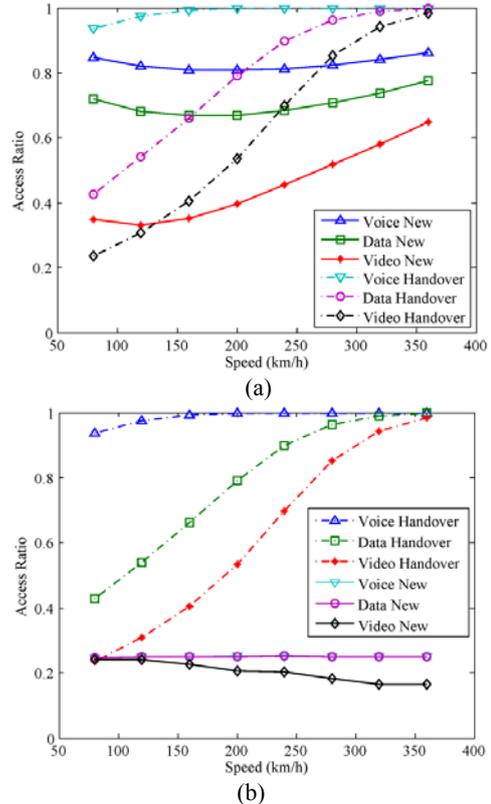

**Fig. 5 Access comparison of handover and new calls**

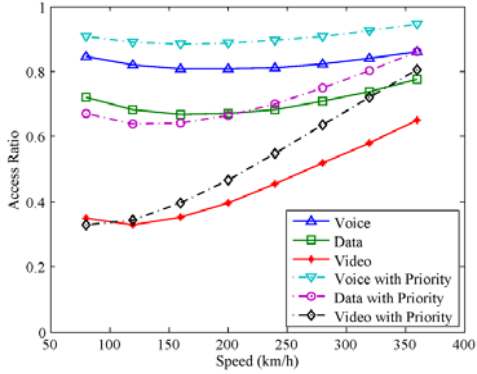

**Fig. 6 Service access considering priority**

Figure 6 shows the access ratios of different applications considering application priority. Note that the voice application priority is the largest, then the video application, and the data application is the least. We can see that the access ratio of the data service with priority becomes lower, while the access ratio of the video service with priority becomes higher compared with the non-priority ones. And the access ratio of the voice service with priority remains the highest. This is obvious because the scheme guarantees the access according to the order of service priority. Hence, our proposed scheme can provide and distinguish the priority requests of different applications for high speed train communications.

Figure 7 shows the access ratios of different applications considering priority and the overhead. With different QoS, the BER performance varies, and the user data is attached with different overhead sizes. Thus, access ratios will decrease compared with the ones that have no overheads. The access ratio of voice application which has the highest priority is less sensitive to the overhead, while the access ratio of data application which has the least priority tends to be impacted the most by the overhead. We can conclude that via considering overhead, the priority order is not changed and a good performance can also be retained. The access ratios performance of in this situation is much similar to the practical situation, which means the proposed scheme can be applied to the real system.

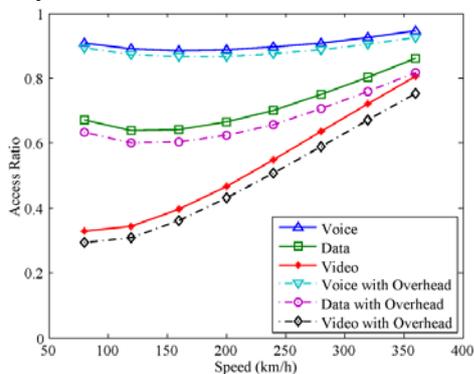

**Fig. 7 Service access considering priority and overhead**

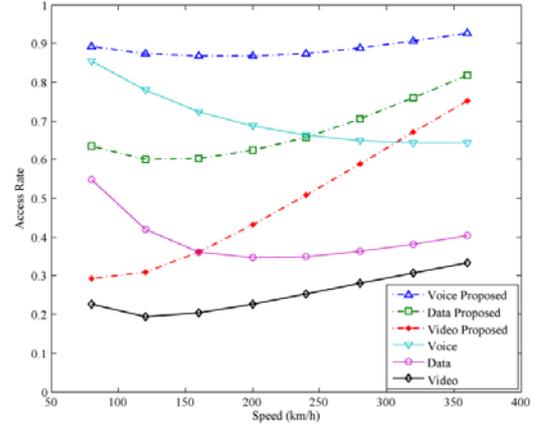

**Fig. 8 Access comparison of models**

In Figure 8, we can see the different performance between the situations of whether MIMO technique is employed in the system or not. Priority and overhead are also considered in these applications. The difference is quite clear that the dropping curves with the increasing speed refer to the system without applying our proposed model and consequently has no power control. Since the required SINR to guarantee the certain QoS cannot be achieved, it shows very poor performance compared with the proposed model, and the gaps are increasing with the mobility becomes higher. Hence, these simulation results show that our proposed model and schemes can provide the QoS requirements of different applications for high speed train communications.

## V. CONCLUSIONS

This paper analyzes the system QoS requirements and MIMO performance for HSR. Two-hop communication model for HSR is applied. A multidimensional resource allocation optimization problem is formulated to satisfy the QoS requirements in the first hop. And dynamic access control schemes are proposed according to different QoS requirements of users in the second hop. Simulation results with comparisons show that the efficiency, reliability, and real-time performances of communication services for HSR users are improved by employing MIMO technologies, resource allocation, and access control schemes. However, some problems such as the detailed BER performance analysis, real-time transmit beamforming realization, and other performance metrics of the proposed access control schemes are not covered in this paper. In addition, many detailed analysis and field tests are needed in order to testify real system performances. The proposed system could be an application or a realization of the future Cyber-Physical Systems (CPS) [26] which argue that all aspects of future system performances are improved to the extent that people expect. We are currently working on these aspects.

**Acknowledgements**

The authors would like to express their great thanks to the support from the 863 Plan of China under Grant 2011AA010104, Beijing Municipal Natural Science Foundation under Grant 4112048, the Key Project of Chinese Ministry of Education under Grant 313006, the Fundamental Research Funds for the Central Universities under Grant 2010JBZ008, and the project of State Key Lab under Grant No. RCS2011ZZ002 and RCS2012ZT013.